\documentclass[12pt]{article}

\usepackage{epsf}
\usepackage{flushrt}
\begin{document}

\title{\bf {\it Chandra} localization of KS~1731--260}

\begin{centering}

\author{\copyright 2001 Ç. M.G. Revnivtsev$^{1,2}$, R.A. Sunyaev$^{2,1}$}

\date{}

\maketitle
{\it $^{1}$ Space Research Institute RAN, Moscow , Russia\\
$^{2}$ MPA, Garching bei Munchen, Germany}

\end{centering}

\vspace{0.3cm}
\begin{center}
{\small Accepted: Aug.7, 2001}\\
{\small To appear in Astr. Lett. 2002, v.1}
\end{center}
\vspace{0.3cm}
\begin{center}
{\bf Asbtract}
\end{center}

We present the analysis of the {\it Chandra} image of KS 1731--260. The 
improvement of the source localization (down to $\sim$0.6'') allowed us 
to rule out 12 of 13 infrared companion candidates proposed by Barret 
et al. (1998). The remaining possible infrared counterpart of KS 1731--260
 (angular distance between sources $\sim1.45$'', or $\sim2\sigma$) has 
the brightness $m_{J}\sim$16 in the J band. If this star is really the counterpart of KS 1731--260, the lower limit on its luminosity is
$L_{tot}>L_{J,H}\sim10L_{\odot}$. The strong drop of the X-ray flux from the source 
allow us to propose the additional check of whether the decribed star is the 
real counterpart of KS 1731--260. If the optical and infrared brightness of 
the binary system is partially caused by the reprocessing of X-rays, as it is
 usual for the low mass binary systems, the infrared brightness of real 
counterpart should strongly reduce in 2001, when the X-ray flux from 
KS~1731--260 turned off (Wijnands et al. 2001).

\vspace{0.6cm}

\section*{}
The X-ray source KS 1731--260 was discovered by TTM telescope aboard MIR-KVANT
observatory in 1988 (Sunyaev 1989, Sunyaev et al. 1990). Since then the source was under 
special attention of the MIR-KVANT-TTM group. During more than 10 years the
source demonstrated moderate X-ray activity at the level of 100-200 mCrab.
KS 1731--260 was observed many times by  GINGA, GRANAT, ROSAT,  ASCA, RXTE 
observatories (e.g. Barret et al. 1998, 2000, Narita 2001). The type I X-ray
bursts were observed from the source, which indicates that the compact 
object in the binary system is a neutron star. The average luminosity of the 
system during its active state in 1989-2000 is of the order of 
$5\times10^{37}$ ergs/s (assuming source distance 7--8 kpc , see Smith et
 al 1997).
At the beginning of 2001 the source flux dropped below the level 
of approximately 10 mCrab and it became undetectable by the RXTE/ASM
 monitor (Wijnands et al 2001). The subsequent observation of {\it Chandra}
observatory revealed a weak
X-ray source with the position consistent with KS1731--260 (Wijnands 
et al. 2001). The source luminosity at that time could be estimated to 
be $L_x\sim2\times10^{33}$ erg/s (Wijnands et al. 2001). The long time 
history of the source according to the MIR-KVANT-TTM and RXTE/ASM data 
could be found in Aleksandrovich et al. (2001).

KS 1731--260 is located within the Galactic plane, which makes its optical 
observations very difficult. The estimation of the photoabsorption from the
 X-ray
measurements give $N_HL\sim1.2\times10^{22}$ cm$^{-2}$ , that corresponds to 
the optical extinction $A_V\sim7.2$  (see e.g. Barret et al. 1998). 
Using the TTM error box Cherepashchuk et al. (1994) proposed two possible 
optical counterparts of KS 1731--260. However, more precise localization of
 the source with ROSAT/HRI ($\sim$10'' error circle) ruled out these 
candidates (Barret et al. 1998). Barret et al. instead
proposed 13 other candidates for infrared counterparts of KS 1731--260. 

In this Letter we present the localization of KS 1731-260 using a {\it Chandra}
observation. Also we discuss the possible source counterpart.

\section*{DATA ANALYSIS AND RESULTS}
In out Letter we used data of {\it Chandra} observation of KS 1731--260,
 performed in Mar. 27, 2001 (the total exposure $\sim$20 ksec). 
The data analysis was done with the help of standard tasks from CIAO 2.1.2. package. The 
spectral analysis  of 
this observation and the discussion of obtained spectral results can be found in Wijnands et al. 2001, Wijnands 2001.

The {\it Chandra} mission requirements on the accuracy of the celestial positions of X-ray sources was 1''. However, the cross-analysis of positions for
known optical and infrared sources with the {\it Chandra} positions shows that
0.6 arcsec could be taken as $rms$ deviation of {\it Chandra}'s values from more
 accurate optical localizations (see
http://asc.harvard.edu/mta/ASPECT/cel\_loc/ cel\_loc.html and
http://asc.harvard.edu/mta/ASPECT/celmon/). Therefore we quote
uncertainties of our localizations as 0.6''.

The {\it Chandra} image gives the source at the position that is consistent with the previous ROSAT/HRI error circle and improve it. The source has the coordinates: RA=17$^h$34$^m$13$^s$.45, Dec=-26$^\circ$05'18''.7(equinox 2000). 
This improvement of the source position allows us to choose between the 
13 possible infrared counterparts proposed in Barret et al. (1998).
In the Fig.1 we present the {\it Chandra} image of KS 1731--260 with the
circles, representing the positions of possible infrared counterparts.
The letters denotes the counterparts, as they were mentioned in Barret 
et al. 1998. It is clearly seen that 12 of 13 possible counterparts 
now can be ruled out. The only candidate, whose position at
 $\sim2\sigma$ level is compatible with the position of KS 1731--260 
is the star H.
(its coordinates from Barret et al. 1998: RA=17$^h$34$^m$13$^s$.35, 
Dec=-26$^\circ$05'18''.0, equinox 2000). However, we should note that
 there is still some ($\sim1.46$'') offset of the candidate counterpart
 from KS 1731--260. The additional infrared observation of this region of 
the sky in needed to confirm the connection of KS 1731--260 with the star H.
If the star H is indeed the counterpart of KS 1731--260 and its optical 
and infrared brightness is partly caused by the reprocessing of the X-rays,
 emitted by the neutron star, as it is the case for most of low mass binary systems,
 we can anticipate that the infrared brightness of the star companion should 
significantly reduce in 2001, when the X-ray source tuned off. If the 
companion in the system is not massive, the change in the optical and infrared
 brightness should be large, however, if the companion is massive the change of
the brightness can be small.

If the star H is indeed the infrared counterpart of KS 1731--260, we can 
estimate its luminosity, or, more precisely, a lower limit on it.
The paper of Barret et al. (1998) give us the brightness of star H: $m_{J}\sim$16 and $m_{H}\sim$15. For a source distance of
7--8 kpc these magnitudes translate into the infrared luminosity of the 
counterpart of 4--5$\times10^{33}$ ergs/s. However, we should take into account 
the optical extinction. The estimates of the source optical extinction 
from the X-ray measurement give us $A_{J}\sim2.0$ É $A_{H}\sim1.25$ (see e.g. Barret et al. 1998). If we take into account these values, the infrared luminosity of the star could be estimated to be 
$L_{J,H}\sim$0.4--1$\times10^{35}$ ergs/s. It means that the total bolometric
star luminosity should be $L>10L_{\odot}$.

{\it Acknowledgements}

This research has made use of data obtained through the {\it Chandra} Data Archive.

\section*{REFERENCES}
Aleksandrovich et al. 2001, Astr.Lett. submitted\\
Barret et al. 1998, A\&A, 329, 965\\
Barret et al. 2000, ApJ, 533, 329
Cherepashchuk et al. 1994, A\&A, 289, 419\\
Narita T., Grindlay  J., Barret D. 2001, ApJ, 547, 420
Smith D, Morgan E., Bradt H., 1997, ApJ, 479, 137L
Sunyaev 1989, IAUC 4839\\
Sunyaev et al. 1990, Sv.Astr.Letters 16, 136\\
Wijnands et al. 2001, ApJ Letters submitted, astro-ph/0107380\\
Wijnands 2001, in Proceedings of Chandra Science Simposium, astro-ph/0107600\\

\begin{figure}
\epsfxsize=14cm
\centerline{\epsfbox{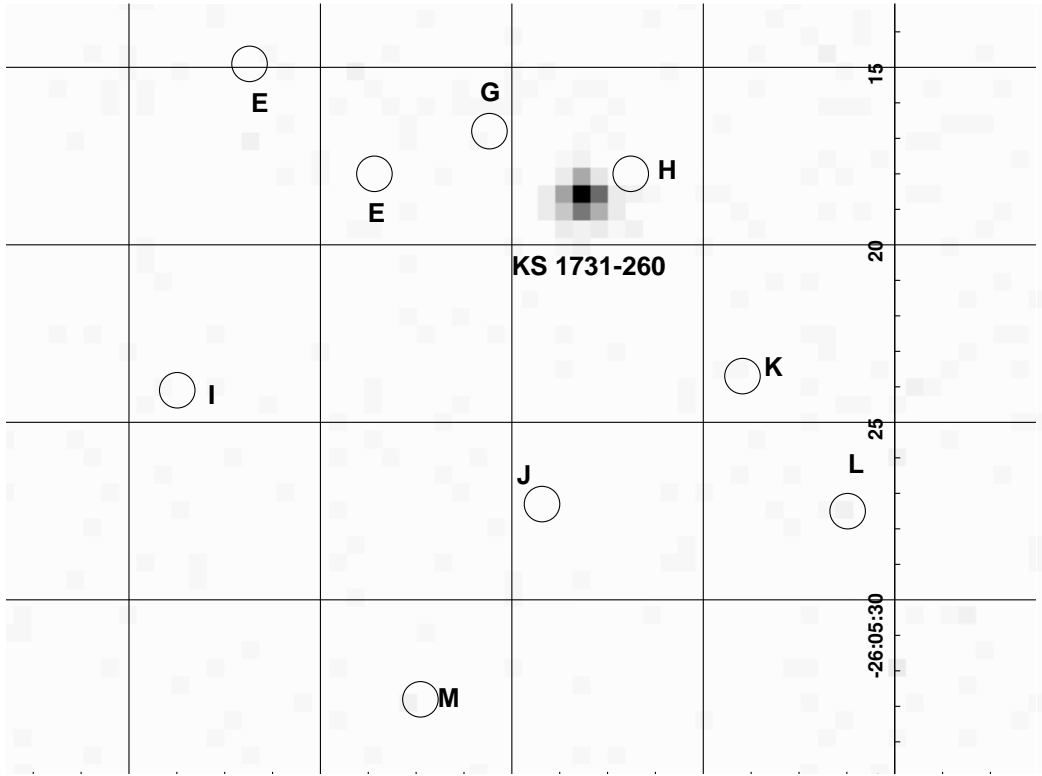}}
\caption{The {\it Chandra} image of the sky region around KS 1731--260. Circles with
letters denotes the positions of the possible infrared counterparts reported in Barret et al. 1998}
\end{figure}

\end{document}